\documentclass[a4paper,11pt]{article}
\usepackage{pos}

\newcommand{\cD}{\ensuremath{\mathcal D} }
\newcommand{\Irm}{\ensuremath{\textrm I} }
\newcommand{\Jrm}{\ensuremath{\textrm J} }
\newcommand{\Krm}{\ensuremath{\textrm K} }
\newcommand{\cN}{\ensuremath{\mathcal N} }
\newcommand{\cO}{\ensuremath{\mathcal O} }
\newcommand{\pq}{\ensuremath{\mathrm{pq}} }
\newcommand{\al}{\ensuremath{\alpha} }
\newcommand{\be}{\ensuremath{\beta} }
\newcommand{\ga}{\ensuremath{\gamma} }
\newcommand{\de}{\ensuremath{\delta} }
\newcommand{\De}{\ensuremath{\Delta} }
\newcommand{\eps}{\ensuremath{\epsilon} }
\newcommand{\la}{\ensuremath{\lambda} }
\newcommand{\lalat}{\ensuremath{\la_{\text{lat}}} }
\newcommand{\mulat}{\ensuremath{\mu_{\text{lat}}} }
\newcommand{\SO}[1]{\ensuremath{\text{SO(}#1\text{)}} }
\newcommand{\Tr}[1]{\ensuremath{\mbox{Tr}\left[ #1 \right]} }
\newcommand{\vev}[1]{\ensuremath{\left\langle #1 \right\rangle} }
\newcommand{\eq}[1]{Eq.~\ref{#1}}
\newcommand{\fig}[1]{Fig.~\ref{#1}}
\newcommand{\secref}[1]{Section~\ref{#1}}
\newcommand{\refcite}[1]{Ref.~\cite{#1}}

\title{Thermal phase structure of dimensionally reduced super-Yang--Mills}

\author*[a]{David Schaich}
\author[b]{Raghav G.~Jha}
\author[c]{Anosh Joseph}

\affiliation[a]{Department of Mathematical Sciences, University of Liverpool, \\ Liverpool L69 7ZL, United Kingdom}
\affiliation[b]{Perimeter Institute for Theoretical Physics, Waterloo, Ontario N2L 2Y5, Canada}
\affiliation[c]{Department of Physical Sciences, Indian Institute of Science Education and Research - Mohali, \\ Knowledge City, Sector 81, SAS Nagar, Punjab 140306, India}

\emailAdd{david.schaich@liverpool.ac.uk}
\emailAdd{raghav.govind.jha@gmail.com}
\emailAdd{anoshjoseph@iisermohali.ac.in}

\FullConference{38th International Symposium on Lattice Field Theory \\ 26--30 July 2021 \\ Zoom/Gather @ Massachusetts Institute of Technology}

\abstract{ 
  We present our current results from ongoing lattice investigations of the Berenstein--Maldacena--Nastase deformation of maximally supersymmetric Yang--Mills quantum mechanics.
  We focus on the thermal phase structure of this theory, which depends on both the temperature $T$ and the deformation parameter $\mu$, through the dimensionless ratios $T / \mu$ and $g = \lambda / \mu^3$ with $\lambda$ the 't~Hooft coupling.
  We determine the critical $T / \mu$ of the confinement transition for couplings $g$ that span three orders of magnitude, to connect weak-coupling perturbative calculations and large-$N$ dual supergravity predictions in the strong-coupling limit.
  Analyzing multiple lattice sizes up to $N_{\tau} = 24$ and numbers of colors up to $N = 16$ allows initial checks of the large-$N$ continuum limit.
}

\begin{document}
\maketitle

\section{Introduction} 
There has been considerable recent interest and progress in the use of lattice field theory to non-pertubatively regularize and analyze supersymmetric quantum field theories.
As reviewed by \refcite{Schaich:2018mmv}, dimensional reduction is a promising means to avoid some of the challenges inherent in lattice supersymmetry, many of which arise from the explicit breaking of the super-Poincar\'e algebra due to the lattice discretization of space-time.
By dimensionally reducing supersymmetric theories all the way down to (0+1)-dimensional quantum mechanics, we end up with significantly simpler systems to analyze.
In addition to reducing the number of degrees of freedom, which makes these theories promising targets for near-term quantum simulation~\cite{Gharibyan:2020bab, Buser:2020cvn, Rinaldi:2021jbg, Culver:2021rxo}, the dimensionally reduced systems also tend to be super-renormalizable.
In many cases a one-loop counterterm calculation suffices to restore supersymmetry in the continuum limit~\cite{Giedt:2004vb}, with no need for the numerical fine-tuning typically required in higher dimensions.

At the same time, these dimensionally reduced theories can often retain interesting non-perturbative dynamics worthy of lattice investigation.
This is certainly the case for the dimensional reduction of maximally supersymmetric Yang--Mills (SYM) theory with $Q = 16$ supercharges and gauge group SU($N$).
The large-$N$ limit of $Q = 16$ SYM quantum mechanics is conjectured to be `holographically' dual to the strong-coupling (M-theory) limit of Type~IIA string theory~\cite{Banks:1996vh, Itzhaki:1998dd, Bergner:2021goh}.
Such SYM quantum mechanics systems consist of balanced collections of interacting bosonic and fermionic $N\times N$ matrices at a single spatial point, and for this reason they are frequently called matrix models.

When considering importance-sampling Monte Carlo analyses of dimensionally reduced SYM, one complication is the presence of flat directions and the resulting integration over a non-compact moduli space, which leaves the thermal partition function ill-defined.
Some deformation of the theory is required in order to lift these flat directions, stabilize numerical calculations, and obtain a well-defined thermal partition function.
A particularly interesting deformation of $Q = 16$ SYM quantum mechanics is the one introduced by Berenstein, Maldacena and Nastase (BMN) in \refcite{Berenstein:2002jq}.
This introduces a dimensionful deformation parameter $\mu$, which can be combined with the dimensionful 't~Hooft coupling $\la = g_{\text{YM}}^2 N$ to define a dimensionless coupling $g \equiv \la / \mu^3$ that maps directly between the lattice and continuum theories.
Similarly, the temperature will appear through the dimensionless ratio $T / \mu$.

The resulting `BMN model' introduces non-zero masses for the nine scalars and sixteen fermions of the theory, explicitly breaking an SO(9) global symmetry (corresponding to the compactified spatial directions of ten-dimensional $\cN = 1$ SYM) down to $\SO{6}\times \SO{3}$.
Notably, it preserves all $Q = 16$ supercharges of the theory, and along with them the holographic connection to M-theory, now formulated on the maximally supersymmetric `pp-wave' background of 11d supergravity that involves plane-fronted waves with parallel rays.
This has allowed numerical construction of dual supergravity solutions~\cite{Costa:2014wya}, which predict the critical $T / \mu$ of a first-order confinement transition in the limit of large $N$ and strong coupling $g$.

In this proceedings we present our current results from ongoing lattice investigations of the thermal phase structure of the BMN model, updating \refcite{Schaich:2020ubh}.
We begin in the next section by summarizing the simple lattice formulation that we use in our numerical calculations, also commenting on discretization artifacts and Pfaffian phase fluctuations.
In \secref{sec:results} we review our numerical calculations and compare our results with weak-coupling perturbation theory and strong-coupling holography.
Finally we conclude in \secref{sec:conc} by briefly reviewing the few steps remaining to finalize our work.

\section{\label{sec:lattice}Lattice formulation} 
As mentioned above, reducing ten-dimensional $\cN = 1$ SYM down to (0+1)-dimensional quantum mechanics introduces an SO(9) global symmetry corresponding to the nine compactified spatial directions.
At the same time, the components of the gauge field in those nine directions become a set of scalars $X_i$ with $i = 1, \cdots, 9$.
The remaining temporal component of the gauge field appears in the covariant derivative $D_{\tau}$.
These bosonic degrees of freedom, and the sixteen fermions $\Psi_{\al}$ with $\al = 1, \cdots, 16$, are all $N \times N$ matrices transforming in the adjoint representation of the SU($N$) gauge group.
A non-zero temperature $T$ is introduced by imposing a finite temporal extent $\be = 1 / T$ with supersymmetry-breaking thermal boundary conditions---periodic for the bosons and anti-periodic for the fermions.
According to holography, at low temperatures with approximate supersymmetry this system corresponds to a black-hole geometry in the dual supergravity.

With the gauge group generators normalized as $\Tr{T^A T^B} = -\de_{AB}$, the continuum action of the BMN model is
\begin{align}
  S       & = S_0 + S_{\mu} \label{eq:action} \\
  S_0     & = \frac{N}{4\la} \int_0^{\be} d\tau \ \mbox{Tr} \Bigg[ -\left(D_{\tau} X_i\right)^2 + \Psi_{\al}^T \ga_{\al\be}^{\tau} D_{\tau} \Psi_{\be}  - \frac{1}{2} \sum_{i < j} \left[X_i, X_j\right]^2 + \frac{1}{\sqrt 2} \Psi_{\al}^T \ga_{\al\be}^i \left[X_i, \Psi_{\be}\right]\Bigg] \cr
  S_{\mu} & = -\frac{N}{4\la} \int_0^{\be} d\tau \ \mbox{Tr} \Bigg[ \left(\frac{\mu}{3} X_{\Irm}\right)^2 + \left(\frac{\mu}{6} X_A\right)^2 + \frac{\mu}{4} \Psi_{\al}^T \ga_{\al \be}^{123} \Psi_{\be} - \frac{\sqrt{2} \mu}{3} \eps_{\Irm\Jrm\Krm} X_{\Irm} X_{\Jrm} X_{\Krm}\Bigg],       \nonumber
\end{align}
where $\ga^{\tau}$ and $\ga^i$ are ten $16\times 16$ Euclidean gamma matrices, with $\ga^{123} \equiv \frac{1}{3!} \eps_{\Irm\Jrm\Krm} \ga^{\Irm} \ga^{\Jrm} \ga^{\Krm}$ and
\begin{equation*} 
  \ga^{\tau} = \begin{pmatrix}0   & I_8 \\
                              I_8 & 0\end{pmatrix}.
\end{equation*}
The $S_0$ in \eq{eq:action} is the dimensionally reduced SYM action, while $S_{\mu}$ is the BMN deformation, in which the indices $\Irm, \Jrm, \Krm = 1, 2, 3$ while $A = 4, \cdots 9$.
As advertised in the introduction, this produces explicit $\SO{9} \to \SO{6}\times \SO{3}$ global symmetry breaking and splits the nine scalars into a set of six $X_A$ and a set of three $X_{\Irm}$.
These two sets of scalars have different $\mu$-dependent masses, and the latter participate in the trilinear interaction known as the `Myers term'.

In our work we employ a simple lattice discretization of this system, replacing the covariant derivative by the nearest-neighbor lattice finite-difference operator
\begin{align}
  \cD_{\tau} & = \begin{pmatrix}0            & \cD_{\tau}^+ \\
                                \cD_{\tau}^- & 0\end{pmatrix} &
  \cD_{\tau}^+ \Psi(n) & = U_{\tau}(n) \Psi(n + 1) U_{\tau}^{\dag}(n) - \Psi(n),
\end{align}
where $\Psi(n)$ are the fermions at lattice site $n$, $U_{\tau}(n)$ is the SU($N$) Wilson gauge link connecting site $n + 1$ (on the right) to $n$ (on the left), and $\cD_{\tau}^-$ is the adjoint of $\cD_{\tau}^+$.
Although more elaborate lattice constructions have been employed by Refs.~\cite{Asano:2018nol, Bergner:2021goh}, it is valuable to compare calculations using a variety of formulations.
Our discretization produces the correct number of fermion degrees of freedom, with no extraneous `doublers'.
However, it does break supersymmetry, for instance by identifying the scalars and fermions with lattice sites, in contrast to the gauge links between neighboring sites.
We retain thermal boundary conditions and a finite temperature, with $\be = 1 / T = aN_{\tau}$ divided into $N_{\tau}$ sites separated by lattice spacing `$a$'.
The dimensionless lattice parameters $\mulat = a\mu$ and $\lalat = a^3\la$ are combined just like the corresponding continuum quantities to continue working in terms of $g = \lalat / \mulat^3$ and $T / \mu = 1 / (N_{\tau} \mulat)$.

\begin{figure}[tbp]
  \centering
  \includegraphics[width=0.45\linewidth]{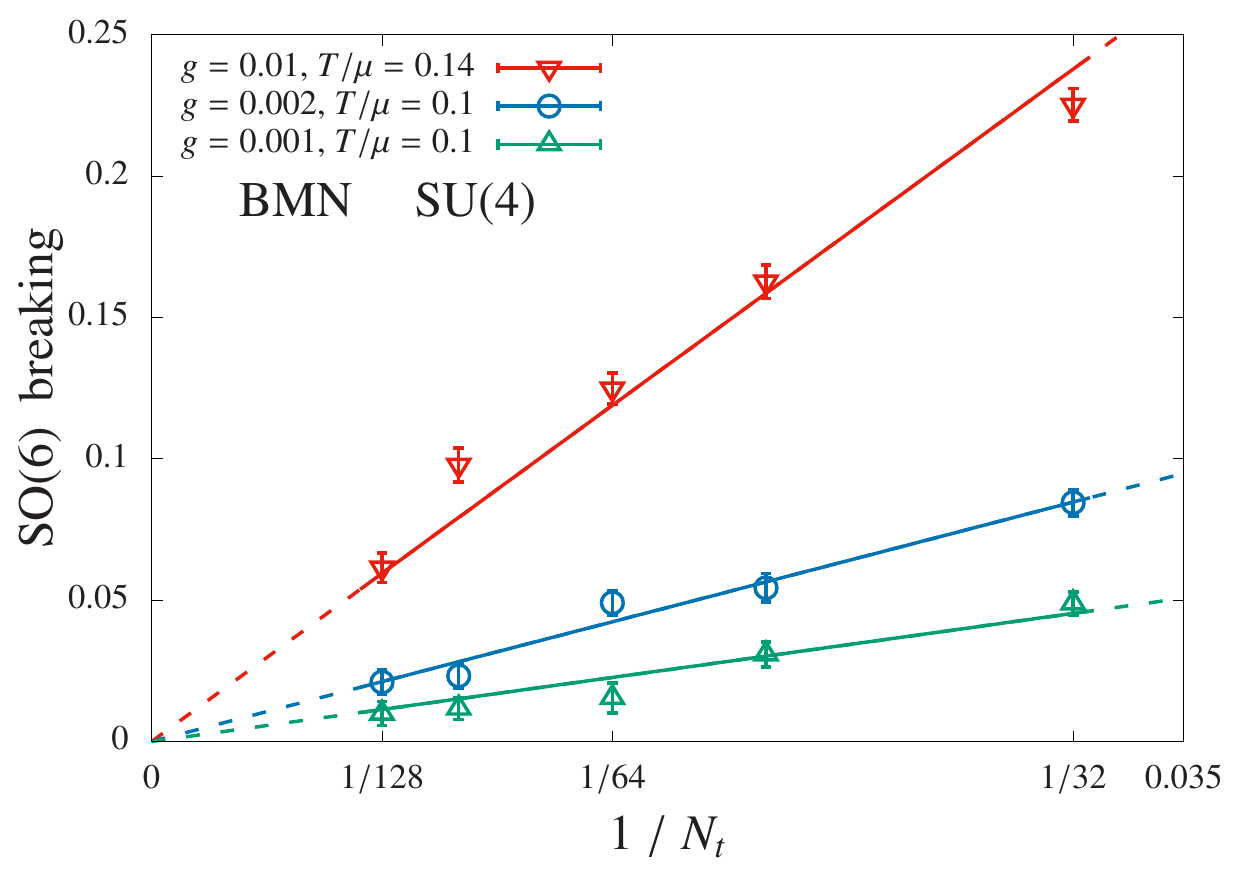}
  \caption{\label{fig:SO6}The breaking of the SO(6) global symmetry quantified by \eq{eq:splitting}, plotted against $1 / N_{\tau}$ for three SU(4) systems with different $g$ and $T / \mu$. In each case the splitting vanishes in the $N_{\tau} \to \infty$ continuum limit, confirming that this is merely a discretization artifact.}
\end{figure}

Our numerical calculations use the standard rational hybrid Monte Carlo (RHMC) algorithm, which we have implemented in the publicly available parallel software package for lattice supersymmetry presented by \refcite{Schaich:2014pda}.\footnote{\texttt{\href{https://github.com/daschaich/susy}{github.com/daschaich/susy}}}
As initially reported in \refcite{Schaich:2020ubh}, our simple lattice action produces apparent discretization artifacts that can be significant for small $N_{\tau}$ at strong couplings $g$.
Specifically, we have observed the six gauge-invariant $\Tr{X_A^2}$ splitting into a set of two with larger values and a set of four with smaller values (but still significantly larger than the three $\Tr{X_{\Irm}^2}$).
This implies a breakdown of the expected SO(6) symmetry, which we quantify by defining the ratio
\begin{equation}
  \label{eq:splitting}
  R_{\text{SO(6)}} \equiv \frac{\vev{\Tr{X_{(2)}^2}} - \vev{\Tr{X_{(4)}^2}}}{\vev{\Tr{X_{(6)}^2}}},
\end{equation}
where $\vev{\Tr{X_{(2)}^2}}$, $\vev{\Tr{X_{(4)}^2}}$ and $\vev{\Tr{X_{(6)}^2}}$ average over the two larger traces, the four smaller traces and all six of them, respectively.
In \fig{fig:SO6} we plot this ratio for three systems with relatively strong $g = 0.001$--$0.01$, considering a small SU(4) gauge group in order to access large lattice sizes up to $N_{\tau} = 128$ close to the continuum limit.
This figure shows that the SO(6) breaking vanishes linearly in the $N_{\tau} \to \infty$ continuum limit, confirming that it is merely a discretization artifact.

\begin{figure}[tbp]
  \includegraphics[width=0.45\linewidth]{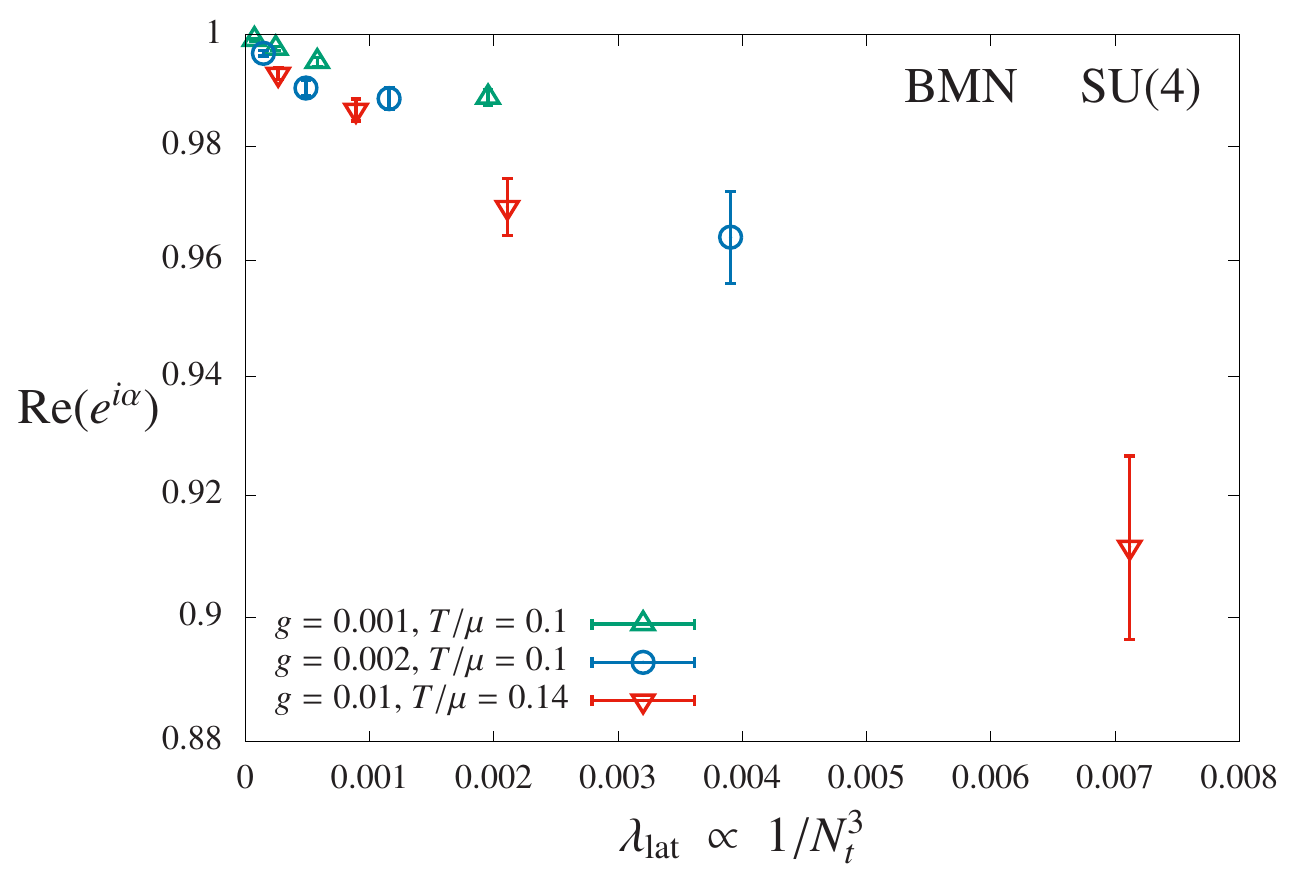}\hfill \includegraphics[width=0.45\linewidth]{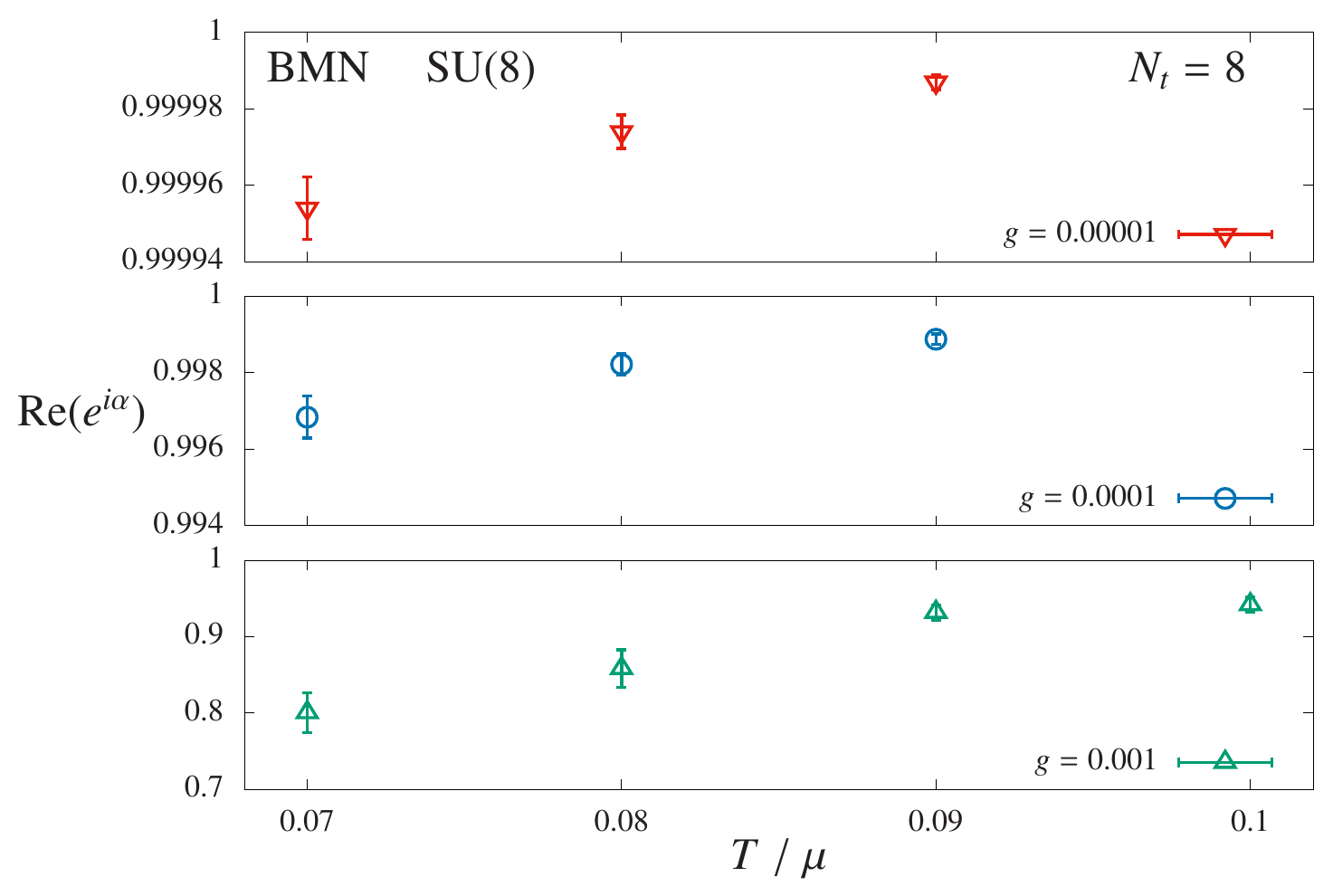}
  \caption{\label{fig:pfaffian}Pfaffian phase fluctuations. \textbf{Left:} Re$(\vev{e^{i\phi}}_{\pq})$ vs.\ $\lalat \propto 1 / N_{\tau}^3$ for gauge group SU(4) confirms that the Pfaffian becomes real and positive in the $N_{\tau} \to \infty$ continuum limit.  \textbf{Right:} Larger fluctuations for larger $g$ with gauge group SU(8) and $N_{\tau} = 8$ are likely related to small-$N_{\tau}$ calculations becoming unstable.}
\end{figure}

Just as for $Q = 16$ SYM in higher dimensions~\cite{Schaich:2018mmv, Sherletov:2021LAT}, integrating over the fermions in the BMN model produces the complex Pfaffian of the fermion operator, which obstructs importance sampling approaches including the RHMC algorithm.
In our calculations we address this issue by `quenching' the Pfaffian phase $e^{i\phi} \to 1$, and need to check whether the phase-quenched (pq) expectation value $\vev{e^{i\phi}}_{\pq} \approx 1$ in order for this to be a good approximation.
Figure~\ref{fig:pfaffian} carries out this check for SU(4) and SU(8) gauge groups, plotting the real part of $\vev{e^{i\phi}}_{\pq}$, which deviates from unity due to fluctuations around the positive real axis.
The left plot in this figure shows these fluctuations vanishing in the $N_{\tau} \to \infty$ continuum limit, thanks to $\lalat = \frac{g}{(T / \mu)^3 N_{\tau}^3} \to 0$ with fixed $g$ and $T / \mu$~\cite{Schaich:2018mmv, Sherletov:2021LAT}.
The right plot fixes $N_{\tau} = 8$ and demonstrates that Pfaffian phase fluctuations increase significantly for stronger couplings $g$, contributing to instabilities we observe in SU(8) $N_{\tau} = 8$ calculations with $g \geq 0.01$.

\section{\label{sec:results}Computational strategy and current results} 
Despite the simplifications offered by investigating the (0+1)-dimensional BMN model, studying the thermal phase structure requires generating a large number of lattice ensembles with the RHMC algorithm.
We adopt the following strategy for these computations.
\begin{itemize}
  \item We need to consider several SU($N$) gauge groups in order to probe the large-$N$ limit where holographic duality is based.
        In this work we compare $N = 8$, $12$ and $16$, in addition to the SU(4) tests of the continuum limit in \fig{fig:SO6}.
  \item For each $N$, we need to consider several lattice sizes $N_{\tau}$ in order to check the $N_{\tau} \to \infty$ continuum limit.
        For this purpose we are using $N_{\tau} = 8$, $16$ and $24$.
  \item For each $\left\{N, N_{\tau}\right\}$, we need to consider a range of dimensionless couplings $g = \lalat / \mulat^3$ reaching from the weak-coupling perturbative regime as close to the strong-coupling holographic regime as we can reach while keeping the numerical calculations stable.
        Overall we investigate four $g = 0.00001$, $0.0001$, $0.001$ and $0.01$, focusing our most expensive SU(16) calculations on the stronger $g = 0.001$ and $0.01$ that are less stable for smaller $\left\{N, N_{\tau}\right\}$.
  \item This procedure leaves us with $25$ distinct $\left\{N, N_{\tau}, g\right\}$, for each of which we need to scan in $T / \mu$ around the confinement transition.
        We begin in the deconfined phase and systematically lower $T / \mu$ through the transition and into the confined phase, initializing each ensemble with a thermalized configuration generated at slightly higher $T / \mu$. 
        Following these initial scans with relatively large $\De (T / \mu) = 0.008$--$0.01$, we carry out refined scans around the transition region with smaller $\De (T / \mu) = 0.001$--$0.002$.
\end{itemize}
Not counting the SU(4) continuum limit study in \fig{fig:SO6}, we have generated $333$ ensembles for our phase diagram study, with up to 50,000 molecular dynamics time units per ensemble.

\begin{figure}[tbp]
  \includegraphics[width=0.45\linewidth]{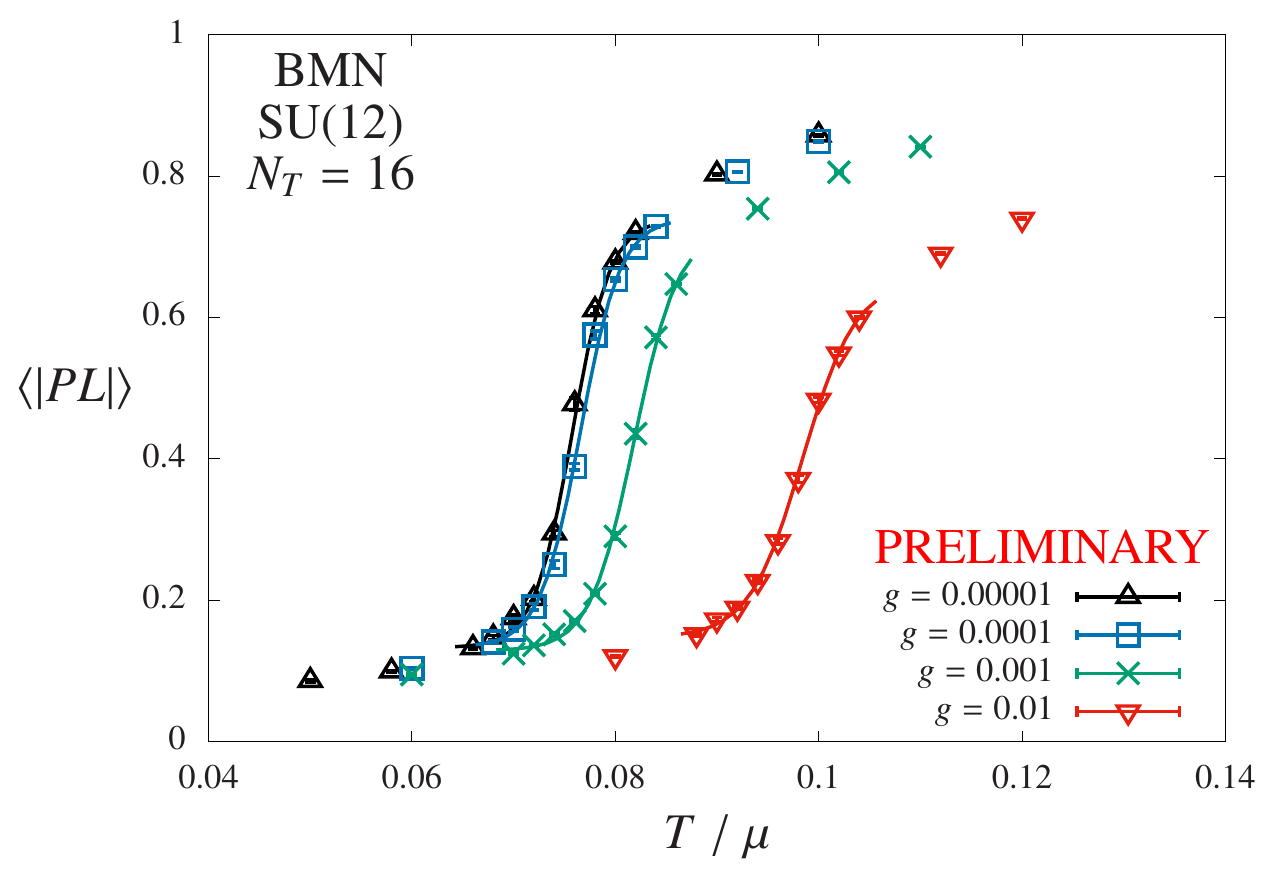}\hfill \includegraphics[width=0.45\linewidth]{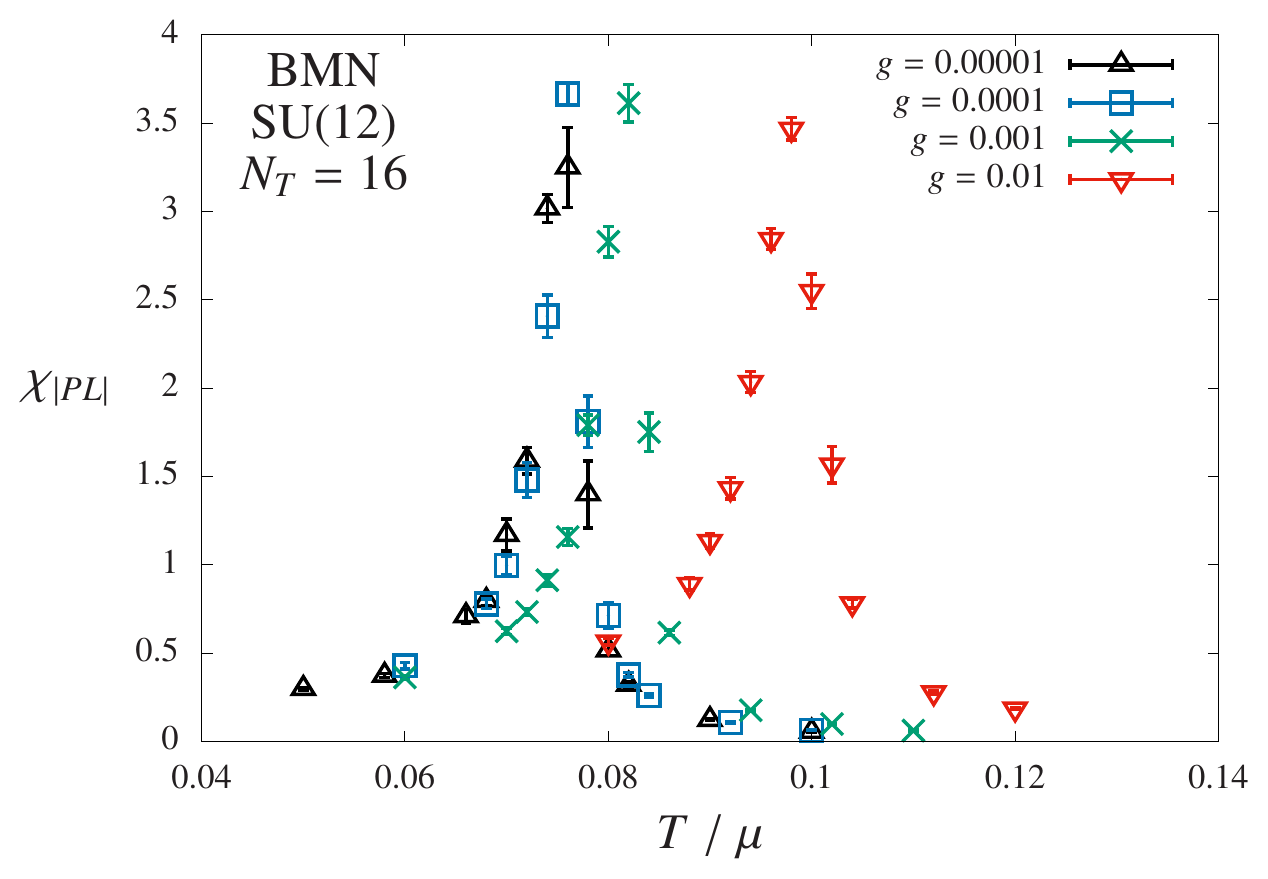}
  \caption{\label{fig:poly}The Polyakov loop magnitude (\textbf{left}) and its susceptibility (\textbf{right}) vs.\ $T / \mu$ for gauge group SU(12) with $N_{\tau} = 16$, considering four $10^{-5} \leq g \leq 10^{-2}$.  The left panel includes preliminary interpolations using a sigmoid ansatz, which predict critical $T / \mu$ values consistent with the peaks in the susceptibilities.  The transition clearly moves to larger $T / \mu$ for stronger couplings $g$.}
\end{figure}

One observable sensitive to the confinement transition of interest is the Polyakov loop, some representative results for which we present in \fig{fig:poly}.
Fixing the gauge group SU(12) and $N_{\tau} = 16$, this figure shows both magnitude of the Polyakov loop, $|PL|$, as well as the corresponding susceptibility $\chi_{|PL|}$ for all four $g = 0.00001$, $0.0001$, $0.001$ and $0.01$.
By interpolating $|PL|$ using a simple four-parameter sigmoid ansatz~\cite{Schaich:2014pda}, we extract values for the critical $T / \mu$ of the confinement transition, which are consistent with the peaks in the susceptibilities.
The transition clearly moves to larger $T / \mu$ for stronger couplings $g$.
While all the ensembles going into this plot have already been generated, the interpolations themselves are preliminary and we continue to experiment with alternate analyses that will inform our estimates of systematic uncertainties on $(T / \mu)_{\text{crit}}$.

\begin{figure}[tbp]
  \includegraphics[width=0.45\linewidth]{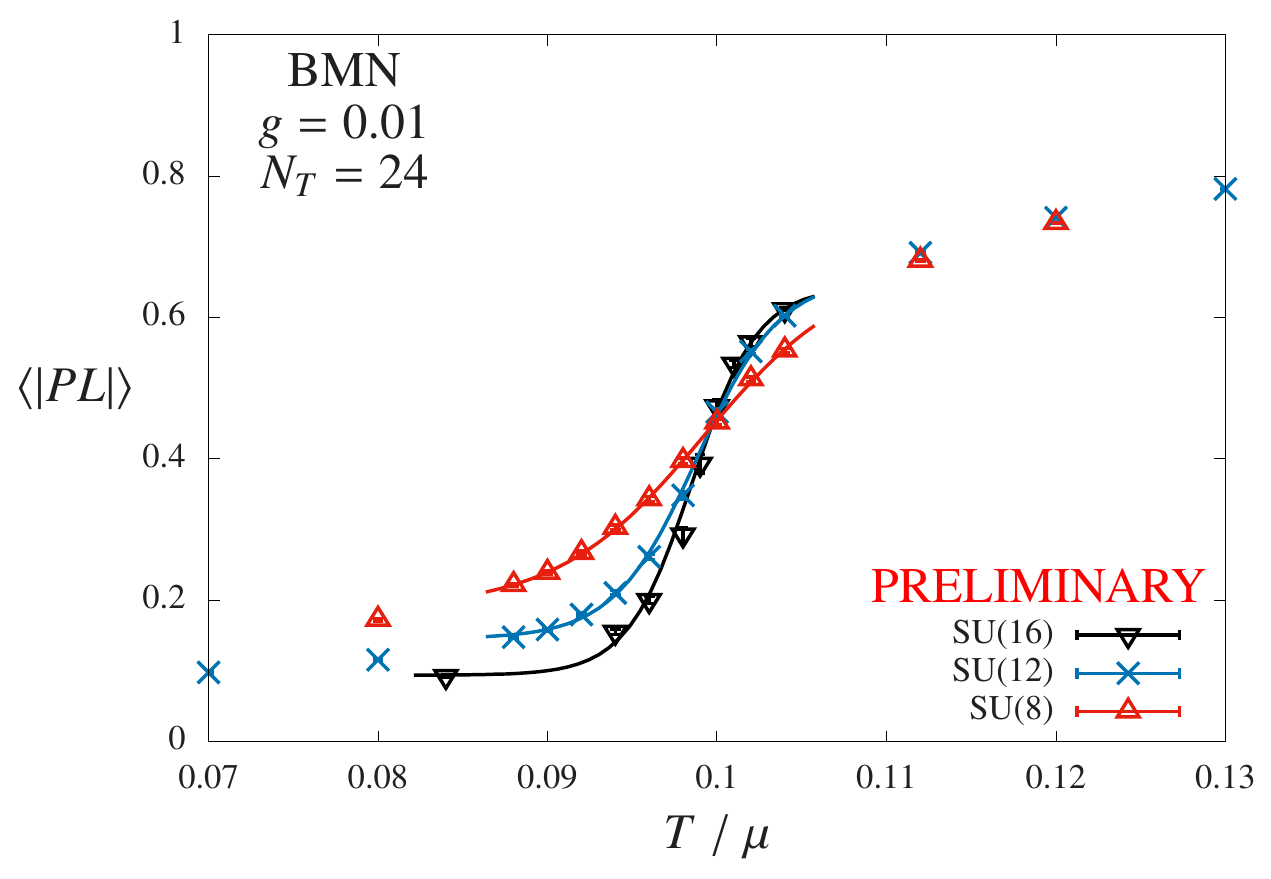}\hfill \includegraphics[width=0.45\linewidth]{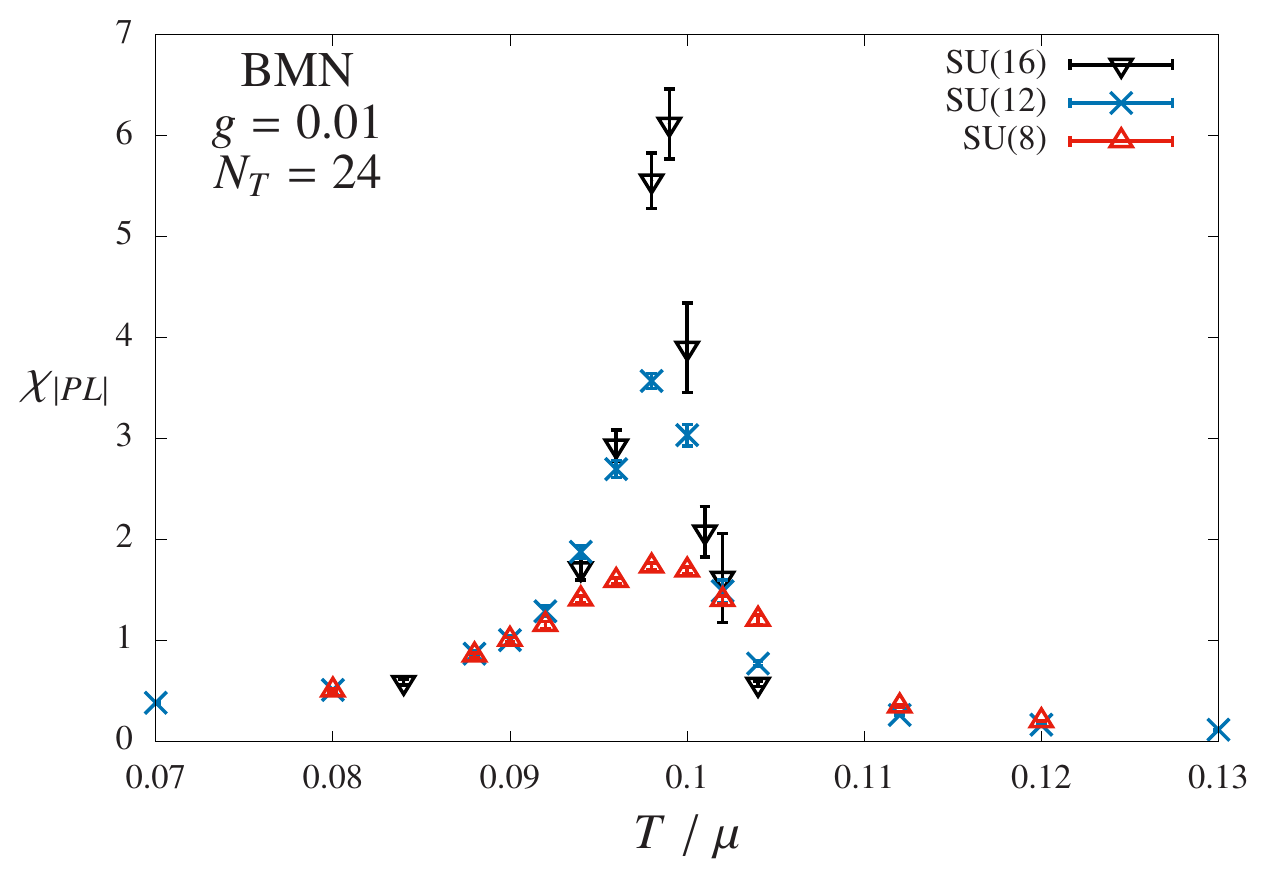}
  \caption{\label{fig:largeN}As in \fig{fig:poly}, the Polyakov loop magnitude (\textbf{left}) and its susceptibility (\textbf{right}) vs.\ $T / \mu$ for $g = 0.01$ with $N_{\tau} = 24$, considering all three SU($N$) gauge groups with $N = 8$, $16$ and $24$. The transition becomes sharper as $N$ increases, with the heights of the susceptibility peaks consistent with a first-order transition.}
\end{figure}

In \fig{fig:largeN} we show a different subset of our Polyakov loop results, again presenting $|PL|$ and the corresponding susceptibility, but now fixing $g = 0.01$ and $N_{\tau} = 24$ while varying the SU($N$) gauge group with $N = 8$, $12$ and $16$.
As expected, the transition becomes sharper as $N$ increases towards the $N^2 \to \infty$ thermodynamic limit.
In particular, the maximum heights of the susceptibility peaks are consistent with the $N^2$ scaling proportional to the number of degrees of freedom that would be expected at a first-order transition.

\begin{figure}[tbp]
  \centering
  \includegraphics[width=0.75\linewidth]{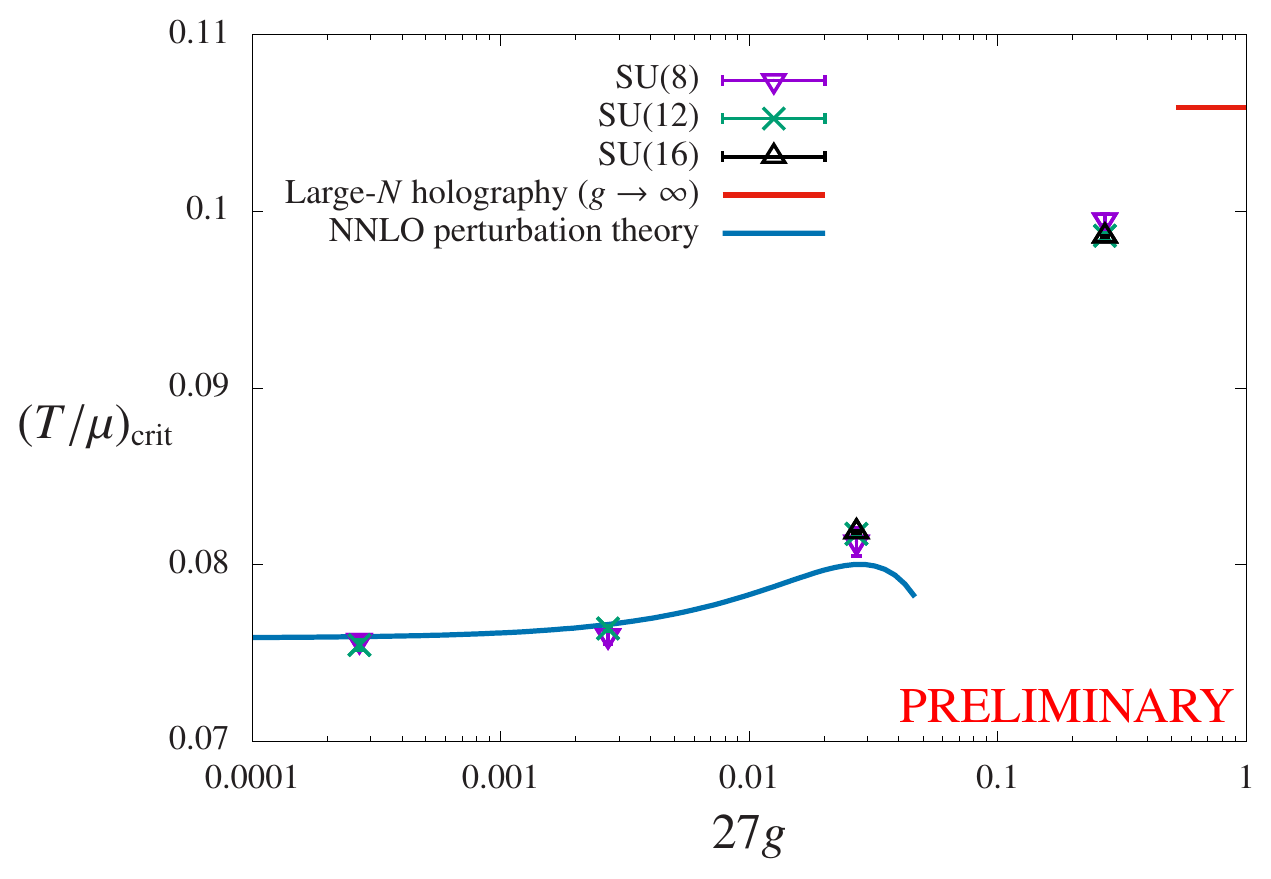}
  \caption{\label{fig:phase}Continuum-extrapolated lattice results for the BMN phase diagram in the plane of critical $T / \mu$ vs.\ the coupling $g$ (times the factor of $27$ appropriate for perturbative expansions).  The lattice results agree with the blue perturbative curve from \eq{eq:NNLO} for sufficiently weak coupling.  As $g$ increases they approach the large-$N$ strong-coupling holographic prediction shown in red. Systematic uncertainties on $(T / \mu)_{\text{crit}}$ remain to be finalized.}
\end{figure}

Such a first-order BMN confinement transition is predicted both by perturbative calculations in the weak-coupling regime as well as by dual supergravity calculations for strong couplings $g = \lalat / \mulat^3 \to \infty$ with $N \to \infty$ and $T / \la^{1/3} = 1 / (N_{\tau} \lalat^{1 / 3}) \ll 1$.
In particular, in the $g \to 0$ limit NNLO perturbation theory (with expansion parameter $27g$) produces~\cite{Furuuchi:2003sy, Spradlin:2004sx, Hadizadeh:2004bf}
\begin{align}
  \label{eq:NNLO}
  \left.\lim_{g \to 0} \frac{T}{\mu}\right|_{\text{crit}} & = \lim_{g \to 0} \frac{1}{12\log 3} \left(1 + \frac{2^6 \cdot 5}{3^4}\left(27 g\right) - C_{\text{NNLO}}\left(27 g\right)^2 + \cO\left(27^3 g^3\right)\right) \approx 0.076 \\
                                                            & C_{\text{NNLO}} = \frac{23\cdot 19\,927}{2^2\cdot 3^7} + \frac{1765769\log 3}{2^4\cdot 3^8}. \nonumber
\end{align}
According to \refcite{Costa:2014wya}, in the strong-coupling holographic regime this critical value should increase to $\left.\lim_{g \to \infty} \frac{T}{\mu}\right|_{\text{crit}} = 0.105905(57)$.
In \fig{fig:phase} we show how our non-perturbative lattice results connect these two regimes, starting off consistent with perturbation theory for $g \lesssim 0.0001$ and then monotonically increasing towards the holographic prediction for stronger couplings.
While the systematic uncertainties on our results for $(T / \mu)_{\text{crit}}$ remain to be finalized, the points in \fig{fig:phase} include extrapolations to the continuum limit, and show no visible dependence on $8 \leq N \leq 16$.

\section{\label{sec:conc}Outlook and next steps} 
In this proceedings we have reviewed our ongoing lattice investigations of the BMN deformation of dimensionally reduced SYM.
In addition to our main focus on the thermal phase structure of this model, we summarized our simple lattice formulation, confirming that SO(6) breaking we have observed is a discretization artifact that vanishes in the continuum limit, where the Pfaffian phase is also real and positive, with no sign problem.
The non-perturbative lattice results we have obtained for the critical $T / \mu$ of the BMN confinement transition are able to connect weak-coupling perturbative expectations and large-$N$ dual supergravity predictions in the strong-coupling limit.
The multiple SU($N$) gauge groups and lattice sizes we have analyzed allow extrapolations to the continuum limit, indicate that the transition is first order, and show no visible dependence of $(T / \mu)_{\text{crit}}$ on $N$.

Although we have completed all RHMC ensemble generation for this project, we continue to work on finalizing our systematic uncertainty estimates for $(T / \mu)_{\text{crit}}$.
In particular, this involves comparing results for $(T / \mu)_{\text{crit}}$ from the sigmoid interpolations shown in Figs.~\ref{fig:poly} and \ref{fig:largeN} with alternate analyses of the Polyakov loop, its susceptibility, and other observables sensitive to the transition.
In addition, we are also analyzing further observables including the distributions of Polyakov loop eigenvalues and the free energy of the dual black hole geometry, the latter of which can be compared with holographic predictions.
More detailed comparisons of our results with those from Refs.~\cite{Asano:2018nol, Bergner:2021goh} may provide insight into the strengths of the various lattice formulations of the BMN model that have been employed so far, providing guidance to future projects.

\vspace{20 pt} 
\noindent \textsc{Acknowledgments:}~We thank Simon Catterall, Toby Wiseman, Denjoe O'Connor, Yuhma Asano, and Masanori Hanada for helpful and interesting exchanges.
RGJ's research at the Perimeter Institute for Theoretical Physics is supported in part by the Government of Canada through the Department of Innovation, Science and Economic Development Canada, and by the Province of Ontario through the Ministry of Colleges and Universities.
AJ was supported in part by the Start-up Research Grant (No.~{SRG/2019/002035}) from the Science and Engineering Research Board, Government of India, and in part by a Seed Grant from the Indian Institute of Science Education and Research (IISER) Mohali.
DS was supported by UK Research and Innovation Future Leader Fellowship {MR/S015418/1} and STFC grant {ST/T000988/1}.
Numerical calculations were carried out at the University of Liverpool and IISER Mohali.

\bibliographystyle{JHEP}
\bibliography{lattice21}
\end{document}